\documentclass{emulateapj}
\usepackage{wrapfig,epsfig,natbib,aas_macros,times}
\usepackage{amssymb}

\newcommand{\Msun}{\mbox{$M_\odot$}}

\usepackage{amssymb}
\usepackage{pdfpages}

\begin{document}

\author{James P. Lloyd}
\affil{Department of Astronomy, Cornell University, Ithaca NY}
\title{The Mass Distribution of Subgiant Planet Hosts}

\date{}



   
\begin{abstract}

High mass stars are hostile to Doppler measurements due to rotation and activity on the main-sequence, so RV searches for planets around massive stars have relied on evolved stars.  A  large number of planets have been found around evolved stars with $M>1.5~\Msun$.
To test the robustness of mass determinations, \citet{Lloyd:2011bh} compared mass distributions of planet hosting subgiants with distributions from integrating isochrones and concluded it is unlikely the subgiant planet hosts are this massive, but rather the mass inferences are systematically in error.
The conclusions of \citet{Lloyd:2011bh} have been called in to question by \citet{Johnson:2013dq}, who show TRILEGAL-based mass distributions disagree with the mass distributions in \citet{Lloyd:2011bh}, which they attribute to  Malmquist bias.   \citet{Johnson:2013dq} argue that 
 the very small spectroscopic observational uncertainties favor high masses, and there are a large number of high mass sub giants in RV surveys.  
However, in this letter, it is shown that Malmquist bias does not impact the mass distributions, but the mass distribution is sensitive to Galaxy model.  The relationship needed to reconcile the subgiant planet host masses with any model of the Galactic stellar population is implausible, and the conclusion of \citet{Lloyd:2011bh} that spectroscopic mass determinations of subgiants are likely to have been overestimated is robust.

\end{abstract}


\maketitle

\begin{figure*}[htbp]
\begin{center}
\includegraphics[width=\textwidth]{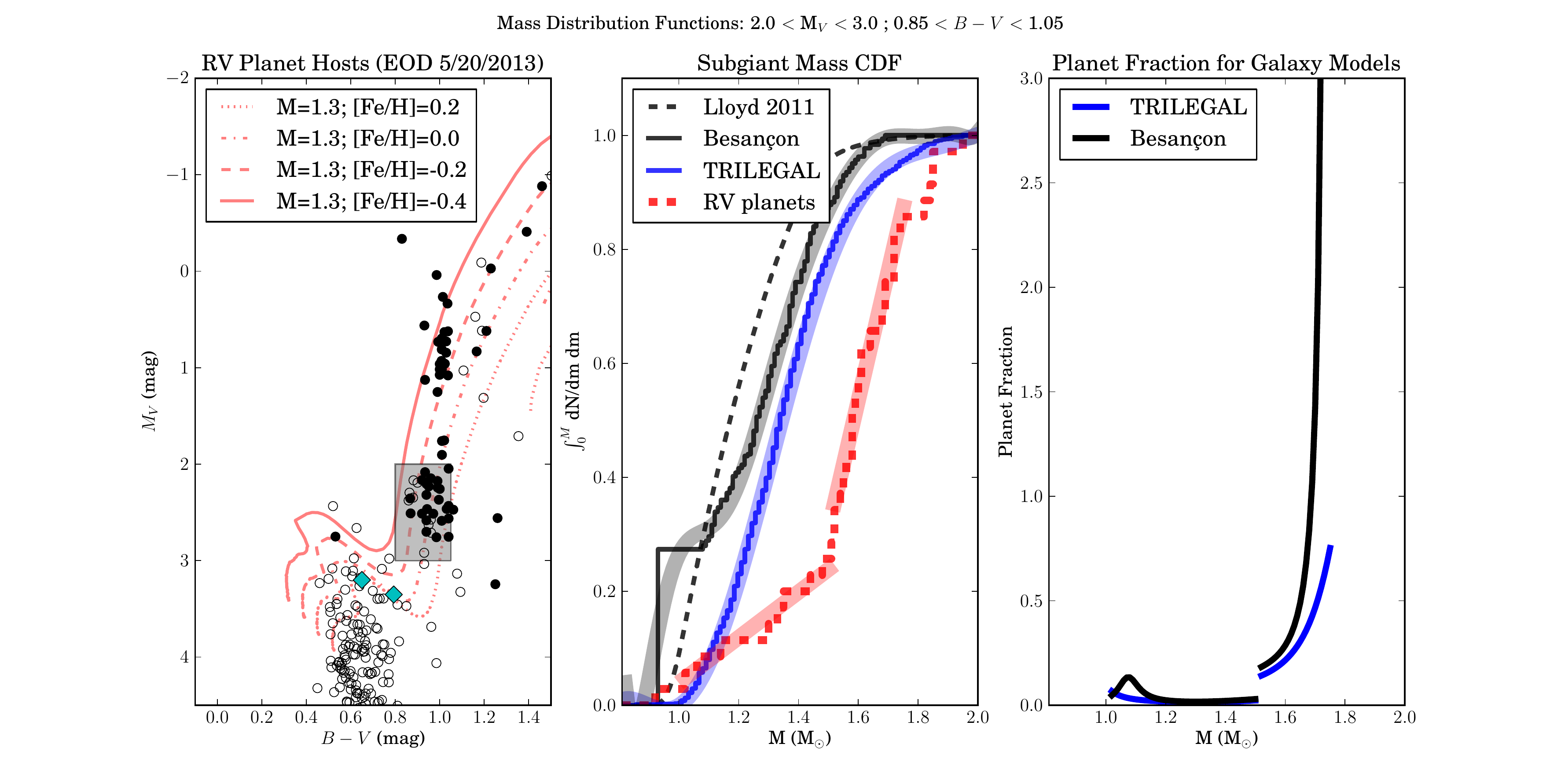}
\caption{
\label{fig:CDF3panel}
	Left : RV planet hosts (5/20/2013 EOD; \citet{Wright:2010rt}).  $M<1.5\Msun$  open symbols,  $M>1.5\Msun$  filled.  The gray shaded region is $2.0<M_V< 3.0;0.8<B-V<1.05$.  Center: cumulative mass distribution functions for $2.0<M_V< 3.0;0.8<B-V<1.05;V<8.5$: Besan\c{c}on  model \citep{Robin:2003fk}, solid black line; TRILEGAL 1.5 \citep{Girardi:2005fk}, solid blue line;  RV planet hosts, dotted red line. The distribution of RV planet hosts is still very different from any galaxy model. Right panel: Planet fraction assuming the 35 subgiant planet hosts come from a sample of 500 stars following exactly the TRILEGAL mass distribution (blue) and Besan\c{c}on mass distribution (black).  The cumulative distributions are interpolated to eliminate binning (polynomial for the galaxy models and piecewise linear for the planets; shown as thick shaded lines in the center panel).  Planet fraction is then calculated from differential counts $(dN/dM)_{\mathrm{planets}}/(dN/dM)_{\mathrm{model}}$.  Planet fraction $>1$ implies that the differential counts at that mass in the 35 planet hosts exceed the differential counts in the model for a sample of 500 stars.
}
\end{center}
\end{figure*}

\section{Introduction}

Planet abundance as a function of stellar mass has shown a sharp increase above $1.5~\Msun$, but with a deficit in short period planets  \citep{Johnson:2007hs,Sato:2008kx,Bowler:2010yq}.  Doppler searches for planets around massive stars have relied on evolved stars because high mass stars are hostile to Doppler measurements  on the main-sequence  due to rotation and activity.  There are now nearly 30 RV planet hosts known as ``retired A-stars'':  subgiants  with  $M>1.5~\Msun$. The conclusion that these stars are  evolved from A-stars is model dependent on  stellar atmosphere and stellar evolution models, both of which have uncertainties for cool stars.  \citet{Lloyd:2011bh} (hereafter L11) demonstrated mass distributions of  subgiants from integrating isochrones are dominated by low masses, and concluded that this population is more likely to have originated from a main-sequence population of lower masses. The deficit of short period planets can be explained by tidal capture, but requires more tidal dissipation than expected.  The planet abundance increase requires either a steep increase in planet frequency with mass, a new migration mechanism, or a high rate of false positives due to signals of stellar origin.  
 
The calculation in L11 used isochrones extending  to the tip of the RGB, and therefore  assumes the subgiants are first-ascent giants.  The arguments here and L11 explicitly do not apply to post-RGB stars.  Many if not all of the planet-hosting stars with $M_V<2$ should be  Helium burning so these arguments should not be  applied to that population.  Clump giants on average younger  than subgiants  \citep{Girardi:2001vn}, so should have a different age/mass distribution.    The situation becomes more complex since mass-loss on the RGB changes the stellar evolution models post-RGB,  and tidal interaction increases with stellar radius star ascends the RGB, which it must do between the subgiant and horizontal branch phases.   The analysis presented here is restricted to the first ascent, ``subgiant'' region.

The conclusions of L11 have been called in to question by \citep{Johnson:2013dq} (hereafter JMW13).   JMW13 find a use the TRILEGAL galaxy model \citep{Girardi:2005fk}, and with a simulated target sample conclude  there is no concern with the abundance of planet hosting sub giants above  $1.5~\Msun$ raised in L11.  JMW13 attributes this difference to
 the inclusion of an apparent magnitude limit (Malmquist bias).
However, as  shown in this letter, even adopting the TRILEGAL model mass distribution in JMW13, the frequency of massive planet hosting subgiants is too large to be attributed to planet occurrence rates.   The difference between the mass distributions calculated in L11 and JMW13 is shown to be a result of the TRILEGAL model parameters, not Malmquist bias.  The mass distribution is sensitive to  Galaxy model, and the Besan\c{c}on Galaxy model \citep{Robin:2003fk} produces a mass distribution with significantly fewer high mass subgiants, insufficient to reconcile with the number of subgiant planet hosts.

\section{Mass Distribution Discrepancy}

JMW13 use spectroscopic  metallicity and effective temperature; Hipparcos  luminosity, combined with YREC stellar evolution models in a Bayesian framework to derive masses.
The mass inferences hinge on  small observational uncertainties  (e.g. the example in JMW13 ascribes uncertainties of: $\pm 50$ K in $T_{\mathrm{Eff}}$; $\pm 0.03$ dex in [Fe/H] (and implicitly exactly solar [$\alpha$/Fe]); $\pm6\%$ in luminosity--3\% uncertainty in distance and negligible uncertainty in bolometric correction).  The SNR and quality of the spectra support  small random errors, but systematic errors could be larger, and have been demonstrated to exist in SME (\citet{Valenti:2005fk}).  \citet{Torres:2012pr} has demonstrated correlations  in SME between Teff and [Fe/H], systematic differences of 100 K  and 0.1 dex in [Fe/H]  between SME and line-by-line analysis, {\em and that these systematic effects lead to errors in mass estimates as large as 15\% for sun-like main-sequence stars}.  Larger systematic differences might be expected in the spectral line fitting for cooler and lower gravity stars than well-calibrated models of the Sun.

Even assuming perfect observations, stellar evolution calculations have uncertainties.  Notably, the treatment of convection in stars is uncertain.  Convective mixing in a stellar core extends main-sequence lifetime and changes the terminal main-sequence core mass.   Quantitatively this uncertainty is especially acute for $1-2\Msun$ \citep{Bressan:2012kx}.  Stellar evolution typically assumes a radiative atmosphere with a gray outer boundary condition, but the formation of molecular hydrogen below 5000K results in energy transport by convection \citep{Chabrier:1997le}.  The impact of the convection parameters on main-sequence stellar evolution is beginning to be explored (cf. Figures 13 and 14 in \cite{Paxton:2013ly}).    Although there are multiple formulations of mixing length theory, in all major stellar evolution databases, it remains the case that  free parameters that are fixed by Solar observational constraints and applied throughout the HR diagram.   In principle there is no reason to expect these parameters to be constant \citep{Salaris:2008fk}.  It would only require a combined offset of 200 K in $T_{\mathrm{Eff}}$ or 0.2 dex in [Fe/H] between observations and  stellar evolution models to shift the mass for a $1.7\Msun$ subgiant to $1.3\Msun$.  Increasing the mixing length $\alpha$ by 5\% is sufficient to introduce such an offset (cf. \citet{Henyey:1965pb};\citet{Pedersen:1990uq}).  

If  subgiants are evolved from lower mass stars, the mass corrections are not necessarily large, but the consequences are profound for our understanding of  tidal evolution, the distribution of planets as a function of stellar mass, and migration mechanisms. \citet{Schlaufman:2013uq}  demonstrated the subgiant planet hosts have kinematic ages (and therefore progenitor masses)  inconsistent with high masses, and concluded tidal evolution is the cause of the deficit of short period planets.   Under this scenario, the abundance of detected planets remains unexplained.

\subsection{Consistency and Disagreement Between L11 and JMW13}

There is agreement between JMW13 and
L11 on the specific question of the general properties of the expected mass distribution in  
subgiant samples.
The prior  in Figure~2 of JMW13
shares   the important properties of the distributions
 in L11: a peak and median less than $1.2\Msun$, a steep
drop between 1.3 and $1.5\Msun$, and a tail beyond $1.5\Msun$. 
%
JMW13 further argue that the precision of  spectroscopic parameters for that star
rules out low mass and the revision to the mass due to 
differential evolution rates is only 3\%.  
JMW13 interpret that the frequency of high mass subgiant planet hosts is purely a due to planet frequency function of stellar mass.  The mass inferences for the Johnson et al. sample without planets are not published, but assuming the TRILEGAL-based subgiant mass distribution in JMW13 requires an unreasonable relationship (see Figure~\ref{fig:CDF3panel}) .

\subsection{Malmquist Bias}

JMW13  argue that the  apparent magnitude 
selection of the sample described in \citet{Johnson:2010ve}
favors high mass subgiants.  However, this is not the case.
 Repeating the calculation in L11 for an apparent magnitude limited
weighting (the calculation is a grid-based integration, not a Monte Carlo sample generation, so is best characterized by ``weighting'', not ``selection'').  The distribution is essentially unchanged (see~Figure~\ref{fig:malmquist}). The scaling argument presented in  JMW13 is based on the separation of evolutionary
tracks in luminosity. However, the selection criteria is  broad  in color ($0.8<B-V< 1.1$), encompassing most of the giant branch. It is true that the
greater luminosity of $M=1.5\Msun$ giants vs $M=1.1\Msun$ giants of a given
metallicity at a given color results in Malmquist bias, 
but the remainder of the lower mass giant branch is still included in the sample.
This is illustrated in Figure~\ref{fig:malmquist}, which shows why the Malmquist bias does not select preferentially by mass.

A second and 
fundamentally important reason that  Malmquist bias does not change  the mass distribution is that the metallicity
spread results in coincidence in the HR diagram of low mass tracks and high mass tracks. For a
single metallicity,  higher mass stars are indeed of higher
luminosity than the lower mass stars. However, also present at the same
position in the HR diagram are lower mass stars of lower metallicity.  A metallicity spread results in low mass stars everywhere on the
giant branch. This is fundamentally a rephrasing of the argument
presented in JMW13: the
color-magnitude selection of stars
produces a mix dominated by low mass stars, and it is only by adopting 
spectroscopically constrained parameters
that high and low masses can be resolved. The sample construction for the radial velocity surveys  \citep{Bowler:2010yq,Johnson:2007hs,Johnson:2008ys,Johnson:2010ve,Johnson:2011qf,Johnson:2011ly}) only
use color-magnitude selections.

\subsection{Scale-Height}

Since Malmquist bias cannot account for the differences between L11 and JMW13, the relevant difference is in models of the Galaxy: homogeneous constant star formation  in the case of L11; a multi-component sophisticated model in JMW13.  The first important difference is the of the scale-height of the disk, the contribution of which can be demonstrated analytically.

Stellar density for an exponential disk with scale-height $z_h$ at  $z$ above/below the mid plane:
$$ \rho_s\propto{e^{-|z|/z_h}}$$
For a mid-plane density $\rho_0$,  surface density integrated over z:
$$\Sigma_0=2 \int_0^\infty{\rho_0}e^{-z/z_h}dz$$
defining $\zeta=-z/z_h$ 
$$\Sigma_0=2\rho_0z_h\int_0^{\infty}e^{-\zeta}d\zeta=2\rho_0 z_h$$

The fraction of  surface density within $\zeta_{\mathrm{max}}=z_{\mathrm{max}}/z_h$ of the mid plane is:
$$\frac{\Sigma'}{\Sigma_0}=\int_0^{\zeta_{\mathrm{max}}}e^{-\zeta}d\zeta=1-e^{-\zeta_{\mathrm{max}}}$$

%
The limiting cases are: constant {\em density} for $z_{\mathrm{max}}{\ll}z_h$, and constant {\em surface density} for $z_{\mathrm{max}}{\gg}z_h$.  However, in the transition between these two regimes, the fraction that are within a volume (apparent magnitude limit for a given absolute magnitude) is {\em exponentially sensitive} to $z_{\mathrm{max}}/z_h$.  For a magnitude limit of $V<8.5$ and luminosity range of $1.8<M_V<3.0$,  the maximum distance is in the range 125-218 pc, comparable to the scale-height.  Since the scale-height increases with age,  the lower mass subgiants are diluted in a sample based on a galaxy model with a strong age-scale-height relation.  

\subsubsection{TRILEGAL scale-height-age relation}
\label{sec:scale-height}

\begin{figure*}
\begin{center}
\includegraphics[width=\textwidth]{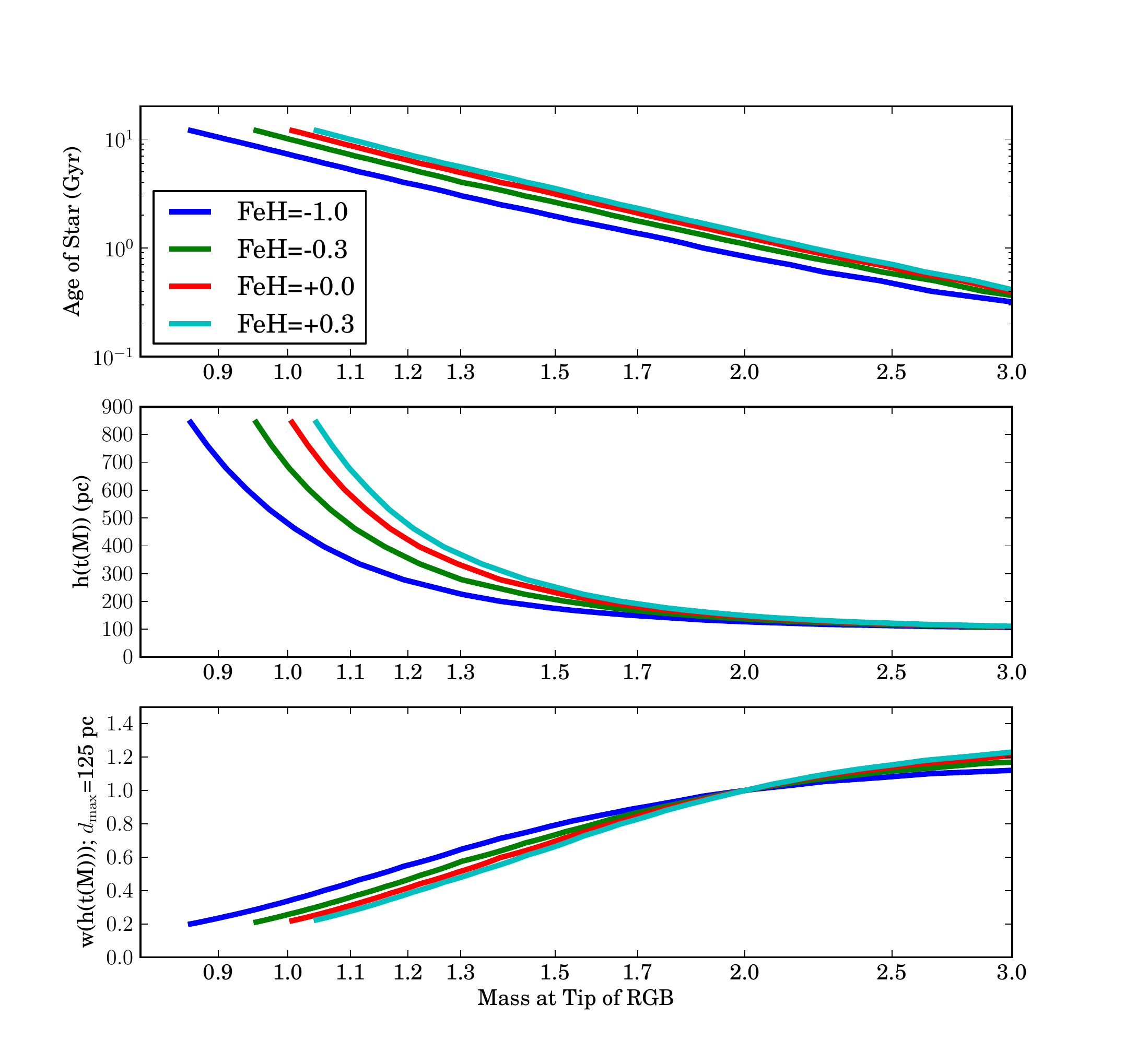}

\caption{Top: Age of the star at the tip of the RGB as a function of stellar mass.  Center: Scale-height vs mass of RGB tip star for TRILEGAL v1.5 parameters and the age-mass relations from the top panel.  Lower panel: weighting of stars of this mass and age in a spherical volume of radius $d_\mathrm{max} =$125, corresponding to a distance modulus of 5.5 mag, for which the $V<8.5$ samples $M_V=3$.  The weighting functions have been normalized at $M=2.0~\Msun$. }
\label{fig:wofhoftofm}
\end{center}
\end{figure*}

\begin{figure*}[htbp]
		\centering
			\includegraphics[width=\textwidth]{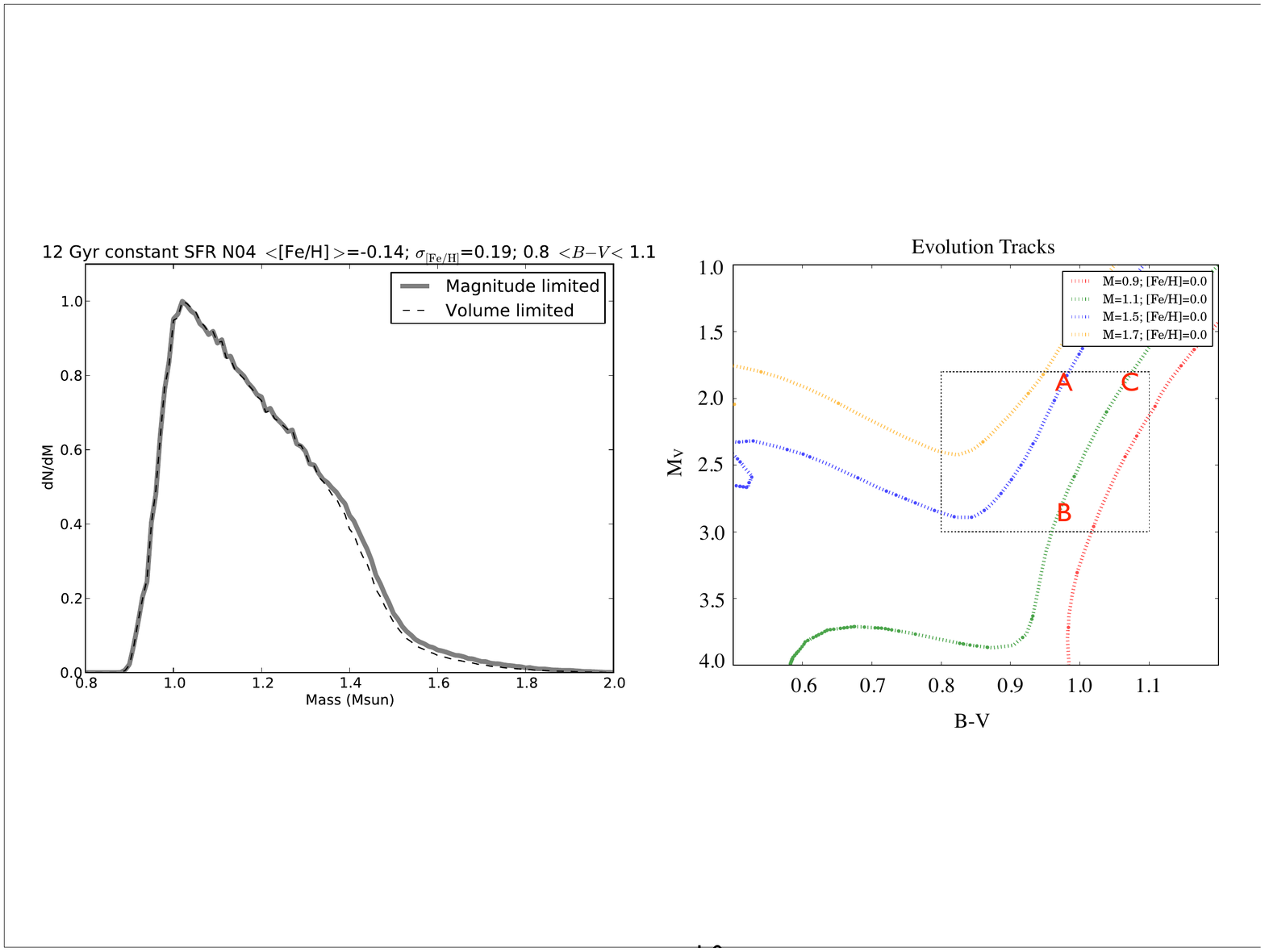}

		\caption{
		\label{fig:malmquist}
	Left, recalculation using the methods of L11, corrected for Malmquist bias.
	Right: Stellar evolution tracks illustrating the impact of Malmquist bias on the sample of stars in the region $0.8<B-V<1.1$;$1.8<M_V<3.0$. JMW13 argue that because higher mass stars are more luminous as subgiants than lower mass stars (cf. the blue $1.5\Msun$  track is well above the green $1.1\Msun$  track), higher mass stars are preferentially included in a magnitude limited sample due to Malmquist bias.  While it is true that high mass subgiants are more luminous (cf. point A is $\sim1$ magnitude above point B), {\em the Malmquist bias applies equally to high mass stars and low mass stars} for a broad color cut (cf. point C is at the same absolute magnitude as point A).		
	 	}
	
	\end{figure*}

TRILEGAL v1.5 uses a  scale-height-age relation:

$$h(t)=z_0\left(1+\frac{t}{t_0}\right)^\alpha$$

with $z_0=95$ pc; $t_0=4.4$ Gyr; $\alpha=1.67$. Relative weighting of different masses due to the age-scale-height relation is shown in Figure~\ref{fig:wofhoftofm}.  This is a genuine astrophysical effect, and although the TRILEGAL relation is steeper at large ages than other work (e.g. \cite{Holmberg:2007kx}; \cite{Robin:2003fk}), this is an important factor in the population synthesis calculations that was not included in L11.  This effect most strongly affects the oldest component,  producing the main difference between the distributions shown in Figure~3 of JMW13: that the L11 distribution peaks near $1.0\Msun$, whereas the JMW13 distribution peaks near $1.3\Msun$.  Since the $M<1.3~\Msun$ subgiant scale-height is larger than the distance limit, these subgiants are preferentially locally diluted relative to higher mass subgiants (Figure~\ref{fig:wofhoftofm}).  This effect does not have a large  impact  on populations with scale-heights less than the distance limit of the sample, so does not account for the differences at higher masses.

%
%
%

%
%

\subsection{Robustness: IMF, SFR and Timescales}

\begin{figure}[htbp]
\begin{center}
\includegraphics[width=\columnwidth]{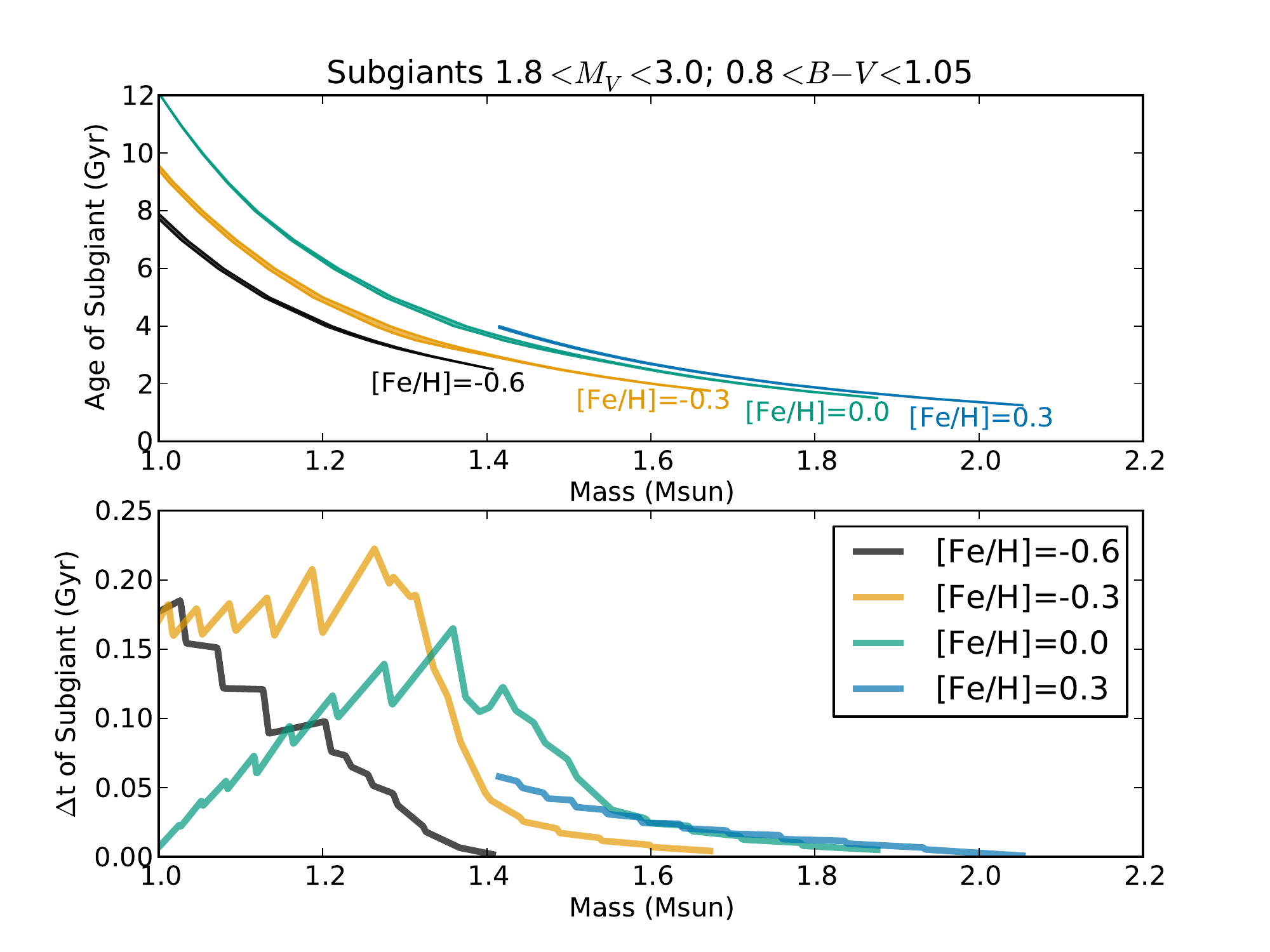}
\caption{Subgiant phase for stars based on the PARSEC isochrones \citep{Bressan:2012kx}.  Top panel: ages of a subgiant star in the region of the HR diagram $1.8<M_V< 3.0;0.8<B-V<1.05$.  Lower panel: time of a subgiant star to transition the subgiant region.  The sharp drop between 1.4 and 1.6 $\Msun$ for all metallicities  is a result of stellar evolution rates.  The slow drop from 1.0 to 1.4 $\Msun$ for [Fe/H]=-0.6 and that there are no $M<1.4\Msun$ subgiants for [Fe/H]=0.3 is a result of the color boundaries.
\label{fig:MassvDt} }
\end{center}
\end{figure}

The argument in L11 depends on three elements.   First, the IMF favors low masses \citep{Salpeter:1955uq,Kroupa:2001qf,Chabrier:2003ve}.  There remain uncertainties in the IMF, but above $1\Msun$ it is well approximated by  Salpeter: $dM/dN=M^{-2.35}$, for which there are 2.25 times as many  $1.0-1.5\Msun$ as $1.5-2.0\Msun$ stars.  There are 3 times as many planet hosting subgiants  $1.5-2.0\Msun$ as $1.0-1.5\Msun$ (Figure~\ref{fig:CDF3panel}).  The ratio between high and low mass subgiant planet hosts is therefore a factor of 7 larger than the ratio of progenitors in the IMF.   Second,  the main-sequence lifetime of  A-stars is $\sim 1.5-2.5$ Gyr,  a brief slice of star formation  in the milky way, whereas lower mass subgiants are sourced from $\sim2-12$ Gyr (Figure~\ref{fig:MassvDt}).  There is agreement star formation rate varies, but has been approximately constant (conservatively within a factor of 2).  Based on consideration of the star formation history that delivers stars to the subgiant region, there should be an additional factor of 10 times fewer high mass subgiants as low mass subgiants.  Finally, the rate at which stars evolve through the subgiant region is a steep function of stellar mass.  Below $\sim1.4~\Msun$, stars leave the main-sequence with degenerate cores and evolve on a shell burning timescale.  Above $\sim1.4~\Msun$, stars leave the main-sequence with thermal pressure supported cores and so evolve more rapidly, on the Kelvin-Helmholtz timescale.  The lifetime of more massive stars within the subgiant region is a small fraction of the lifetime of lower mass stars (Figure~\ref{fig:MassvDt}).  There is yet another order of magnitude suppression in the fraction of higher mass stars subgiants due to differential evolution rates.  Based only on these considerations, it is surprising that there are 27 subgiant stars above $1.5\Msun$, let alone that 27 planets have already been discovered with host masses above $1.5\Msun$ in the sample of 500 stars defined by \citet{Johnson:2010ve}.

\begin{figure*}[htbp]
\begin{center}
\includegraphics[width=\textwidth]{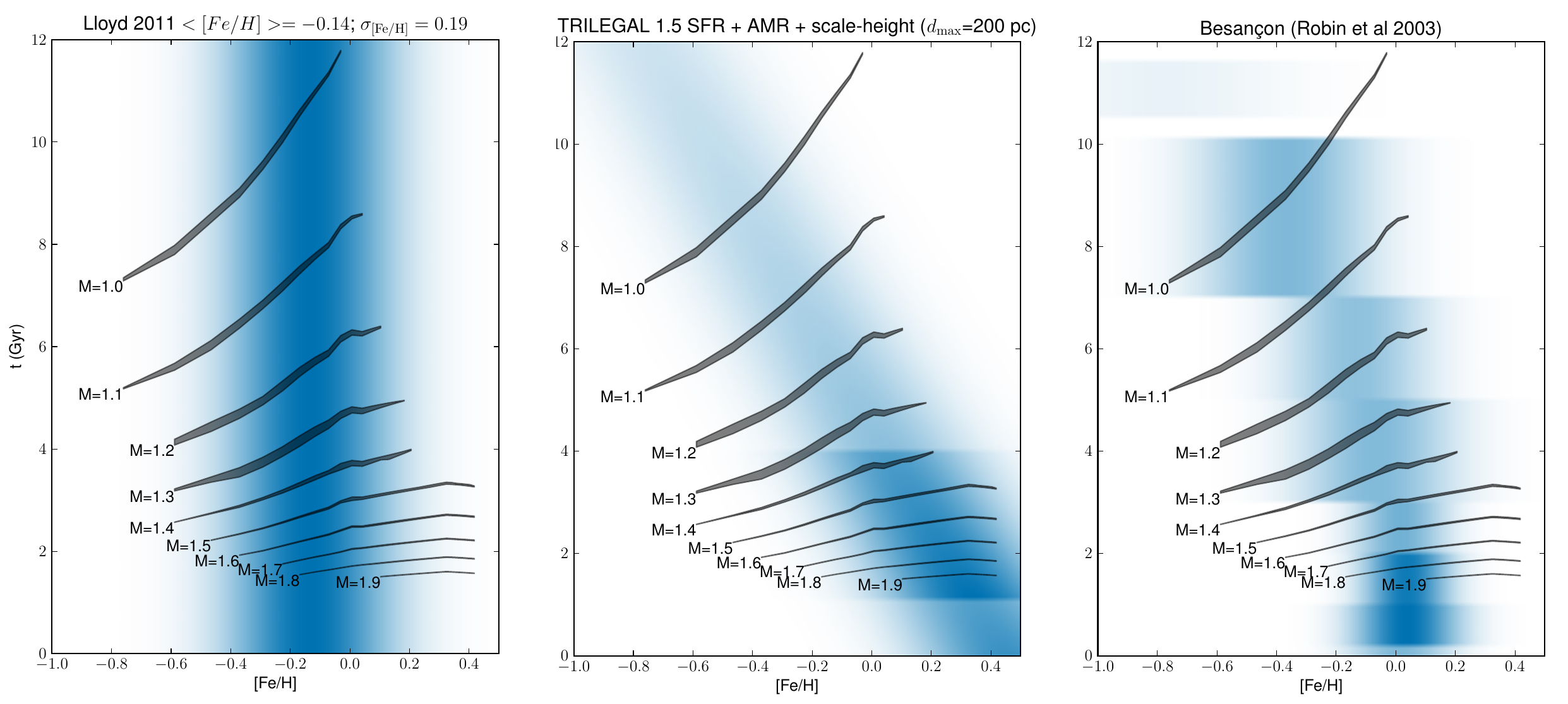}
\caption{Galaxy models in metallicity-age.  Blue shows density of stars in the solar neighborhood of a given metallicity and age for each Galaxy model.  Gray shaded regions show the locus of age-metallicities that result in a star   in the subgiant star region: $1.8<M_V<3.0;0.8<B-V<1.05$ (same methods as Figure~\ref{fig:MassvDt}).  Each Galaxy model has differing assumptions, resulting in a different mass distribution of the subgiants (cf. Figure~\ref{fig:CDF3panel}).
The simplistic L11 model assumes a constant star formation history for 12 Gyr, with a metallicity spread \citep{Nordstrom:2004fj}.  The TRILEGAL 1.5 model \citep{Girardi:2005fk} includes effects that produce more higher mass stars in the subgiant region: 1)  age-metallicity relation; 2) 50\% increase in the star formation rate from 1-4 Gyr; 3) scale-height vs age relation that dilutes the solar neighborhood density for old stars (see Section~\ref{sec:scale-height}).  The Besan\c{c}on Galaxy model \citep{Robin:2003fk}, has less pronounced versions  these effects, and produces a mass distribution intermediate between  L11 and JMW13 (cf. Figure~\ref{fig:CDF3panel}).
\label{fig:GalaxyModels}
}
\end{center}
\end{figure*}

There are factors that  skew the distribution of subgiants to higher masses: the effect of an age-scale-height relation;  star formation rate as a function of time (e.g.  enhancement  from 1-4 Gyr  in the TRILEGAL model);  age-metallicity relation, which slows evolution  and expands the epoch of star formation history feeding the subgiant region.  The impact of the Galaxy model assumptions on the mass distributions is shown in Figures~\ref{fig:CDF3panel}~and~\ref{fig:GalaxyModels}. Dependance in the distribution of planets on stellar mass is to be expected (e.g. \citet{Fressin:2013cq,Howard:2013hq,Ida:2005zv,Johnson:2007hs,Laughlin:2004cr,Mordasini:2012yo,Spiegel:2012qc,Udry:2007fh,Vigan:2012fc}).   Age-metallicity relations and stellar evolution rates also interact with planet-metallicity correlation \citep{Fischer:2005mz} in 
ways that should alter the abundance of planets orbiting subgiants.  
However, the relationship needed to reconcile the subgiant planet hosts with all models of the Galactic stellar population is  implausible, and inconsistent at the lower mass end, where there is overlap with RV searches targeting FG dwarf stars (cf. the low absolute value and decline in relative planet fraction between 1.0 and 1.4~$\Msun$ and the drastic rise at $1.5\Msun$ in the right panel of Figure~\ref{fig:CDF3panel}).   Of the 35 planet hosting subgiants, 9 are M$<1.5\Msun$, 21 between 1.5 and $1.75\Msun$ and 5 above $1.75\Msun$.    Assuming these  were drawn from a sample of 500 subgiants distributed according to the TRILEGAL model, there should be 385, 94 and 21 subgiants in these three bins, resulting in detected planet frequencies of 2.2\%, 24\% and 21\% respectively.  For a sample  distributed according to the Besan\c{c}on model, there should be 433, 64 and 3 subgiants in these three bins, resulting in detected planet frequencies of 2.0\%, 38\% and 160\% respectively.   There is a sharp rise in $dN/dM$ of the planet hosts at $M=1.5\Msun$, at which $dN/dM$ is continuous and decreasing for all Galaxy models, so it is unavoidable to have a jump in the planet frequency of more than a factor of 10 at this boundary if the spectroscopic stellar masses are correct. Although there are sufficient $M>1.5~\Msun$ subgiants in the TRILEGAL model to account for the RV planet hosts after allowing for a such a relation between planet frequency stellar mass, this relation is so extreme as to appear implausible.  The Besan\c{c}on model of the Galaxy contains fewer subgiants above $1.68\Msun$ in a subgiant sample of 500 stars than the number of subgiant planet hosts above $1.68\Msun$ in the Exoplanet Orbit  Database.

The simplest interpretation is that the spectroscopically inferred masses (the method described in JMW13) has led to systematic over-estimation of  stellar masses.  Although stellar evolution and spectral synthesis are generally robust and  well tested, the mass inferences depend on very high precision and an assumption that there are no systematic offsets between the stellar evolution scales and the observational scales.  

%
%

\section{Discusssion}

The distribution of RV semi-amplitude ($K$) for the subgiant planets is remarkably different to the distribution for either  main-sequence stars or  more evolved giants \citep{Schlaufman:2013uq}, in particular deficient in large amplitudes.  That the largest amplitude signals of $K\sim50$ m/s are only a factor of a few larger than lowest amplitude signals published as confident detections implies that all the observed signals are only of moderate confidence, and is suggestive that false-positive signals could be at play.  There is no explicit evidence to suggest false positives, but it is worth noting that the subgiant planet signals have not been subjected to the same level of scrutiny as the solar-type stars (e.g. \citet{Gray:1997fj,Gray:1997uq,Gray:1998kx,Marcy:1998zr,Walker:1997mz,Willems:1997ly}), nor has the planetary interpretation been confirmed by the presence of transits as it has for solar-type stars \citep{Charbonneau:2000ve}.  These long periods are comparable to the rotation or convective overturn timescale of the stars, and reliance on the case made for the solidity of the planetary interpretation of Doppler signals made for solar-type stars should be treated with caution when extrapolated.  If not false positives, then the abundance increase requires another explanation.   \citet{Schlaufman:2013uq} have suggested planet-metallicity effects and a metal-rich subgiant population, but this requires the observations have determined systematically low spectroscopic metallicities for the subgiants, which would further exacerbate the mass distribution problem, and does not explain the $K$ distribution.

An increase in the abundance of planets around subgiants could be explained by migration of planets from an outer reservoir at long periods, assuming that planets like Jupiter and Saturn are common.  Stellar tidal migration timescales are typically very long \citep{Rasio:1996qf}.  Although there is large uncertainty in tidal dissipation rates, conventional tidal theory is challenged by the interpretation that planets within 0.5 AU will be engulfed by subgiants \citep{Hansen:2010fy,Hansen:2012bh}, and stellar tides could not migrate planets from several AU.  A speculative migration mechanism would be a gravitational torque induced by the planet interacting with  stellar wind.  \citet{Cranmer:2011sh} have shown that the stellar wind (as might be expected to explain the rotational evolution of subgiant stars) may involve significant mass-loss on the subgiant branch, more than would be predicted by application of the \citet{Reimers:1975xd} mass-loss relation to the subgiants.   In the absence of interaction between the planet and  stellar wind, mass-loss induces outward migration due to the decrease in stellar mass \citep{Sackmann:1993dq}.  However, the planet can tidally distort the stellar wind and the back-reaction torque of the distorted wind could induce migration  analogous to the migration mechanisms in protoplanetary disks \citep{Lin:1986cr}.  If the  mass-loss rate is low, the instantaneous mass in the wind, and therefore migration rate will be low, but the total migration will depend on the migration rate $\times$ the duration, and is therefore sensitive to the {\em total mass-loss}, not the mass-loss rates.   In principle, mass loss of $10^{-3}$ to $10^{-2}\Msun$ over $10^8$ yr could be sufficient to migrate a inwards planet by gravitational torque. 

Revision of the stellar masses of the subgiant planet hosts to masses comparable to the  dwarf stars in radial velocity surveys presents a serious challenge to explain the dramatic increase in planet abundance.  If this can no longer be attributed  purely to a difference in stellar mass, then it needs another explanation.    False positive signals can explain the abundance increase and the $K$ distribution, but there is no evidence for his interpretation.  If there is a common mechanism that migrates Jupiter-like planets from Jupiter-like distances inwards to orbits of $\sim 1$ AU during the early sub-giant phase of stellar evolution, this could explain the observations without false positives.  This scenario would profoundly change our view of the future of the Earth, which would be most likely be destroyed if subjected to a Jupiter-crossing orbit.

\section{Acknowledgements}

The author is indebted to John Johnson, Tim Morton, Jason Wright, Adam Burrows, Kevin Schlaufman, Josh Winn,  and David Chernoff for conversations that influenced this work.


\end{document}